\RequirePackage[2020-02-02]{latexrelease}
\documentclass[prl,showpacs,preprintnumbers,amsmath,amssymb]{revtex4}
\usepackage[dvips]{graphics}
\usepackage{tikz}
\begin{document}

\title{Planar Rotor in Matrix Mechanics and the Role of States in Quantum Physics}

\author{Vlatko Vedral}
\affiliation{Clarendon Laboratory, University of Oxford, Parks Road, Oxford OX1 3PU, United Kingdom}

\date{\today}

\begin{abstract}
\noindent We illustrate Heisenberg's method of matrix mechanics using the planar quantum rotor example. We show how to find the spectrum of this simple model without the need to use the eigenstates of the system. This then leads us to speculate on the role the Heisenberg state plays in quantum mechanics and to ask whether one could even completely dispose of the need for states.  
\end{abstract}

\pacs{03.67.Mn, 03.65.Ud}% PACS, the Physics and Astronomy

\maketitle                           %display desi d

The main issue we would like to explore in this note is the role of states in quantum physics. Specifically, could we do without them? The reason why this question is even meaningful is that the first paper of Heisenberg's on quantum mechanics does exactly that \cite{Longair,Born,Green,Waerden}. He finds the spectra of harmonic and anharmonic oscillators without ever finding the corresponding eigenstates. Of course, this was exactly the information needed to model the existing experiments on atomic spectra. Surprisingly, however, we will show that the knowledge of the eigenstates is in general not required. 

Here we illustrate Heisenberg's method on an even simple system: the planar rotor, or, as it is also known, a particle on a ring. We first show how to obtain its spectrum without ever needing the information about the eigenstates. An analysis of the rotor was present in the original paper of Heisenberg's on quantum mechanics and, subsequently, the full treatment of angular momentum was given by Born, Heisenberg and Jordan \cite{Waerden,Frenkel}. Then, we discuss if the states are needed at all and how we could either minimize their ontological relevance by simply downgrading them to an initial condition. 

The Hamiltonian for a rotor is given by $H=L^2/2I$ where $L$ is the angular momentum (perpendicular to the plane of rotation) and $I=\mu r^2$ is the moment of inertia of the particles of mass $\mu$ revolving around the ring of radius $r$. 

First of all, it is clear that diagonalizing this Hamiltonian is tantamount to diagonalizing the operator $L$. Therefore the $n$-th energy eigenvalue is given by $E_n = L^2_{nn}/2I$ where $L_{nn}$ is the $n$-th diagonal element of the matrix representing $L$. It is given by $L_{nn} = \langle n|L|n\rangle$, where $|n\rangle$ is the $n$-th eigenstate, but, as we said, the trick is to find $L_{nn}$ without having to find an explicit form for $|n\rangle$. 

We next proceed to set up the commutation relations between conjugate variables. Normally, one thinks of the angle $\phi$ as being conjugate to $L$. However, the problem then is that the commutator $[\phi,L]=i\hbar$ is impossible to implement \cite{Louisell,Carruthers}. The reason is that the ranges of the two observables are different, since $\phi$ varies between $0$ and $2\pi$ and $L$ between $-\infty$ and $+\infty$. The commutator leads to the uncertainty relations of the kind $\Delta \phi \Delta L \geq \hbar$, however, if $\Delta L \rightarrow 0$, this implies that $\Delta \phi \rightarrow \infty$ which is clearly impossible as it exceeds the value of $2\pi$. 

It turns out that the appropriate commutation relations are: 
\begin{eqnarray}
\bigl[L,\sin\phi\bigr] & = & -i\cos \phi \\
\bigl[L,\cos\phi\bigr] & = & i\sin \phi
\end{eqnarray}
where now the entities $\sin\phi$ and $\cos\phi$ are treated as quantum mechanical operators. We proceed to calculate the $nm$ elements of these equalities. Looking at the first commutator
\begin{equation}
L_{nn}(\sin\phi)_{nm} - (\sin\phi)_{nm} L_{mm} = - i\hbar (\cos \phi)_{nm}
\end{equation}
since $L$'s only non-zero elements are the diagonal ones. The formula resulting from the second commutator is similar. 

In order to calculate $(\sin\phi)_{nm}$ and $(\cos \phi)_{nm}$, we exploit the fact that, classically, $x=r\cos\phi$ and $y=r\sin\phi$. In terms of the Heisenberg's matrix description, we have that 
\begin{eqnarray}
x_{nm} & = & r (\cos \phi)_{nm} e^{\omega_{nm}t} \\
y_{nm} & = & r (\sin \phi)_{nm} e^{\omega_{nm}t}
\end{eqnarray}
In other words, both $x$ and $y$ coordinates undergo a harmonic motion at the frequency $\omega$ such that $\phi = \omega t$. If this is to be consistent with the operator equations, we need to have that
\begin{eqnarray}
(\omega^2_{nm}-\omega) x_{nm} & = & 0 \\
(\omega^2_{nm}-\omega) y_{nm} & = & 0
\end{eqnarray}
This implies that the only time the elements $x_{nm}$ and $y_{nm}$ are non-zero is when $\omega_{nm}=\pm \omega$. Note that the quantum-classical correspondence is at the heart of this kind of reasoning. The quantum equations of motion must look the same as the corresponding classical ones; the crucial difference is that the quantum equations are obeyed by operators while classical ones are obeyed by numbers. 

We are still at liberty to choose the elements of our operators freely and it is customary to do so in the way that the non-zero elements are $x_{nn+1}$ (corresponding to $\omega_{nn+1}=\omega$) and $x_{nn-1}$ (corresponding to $\omega_{nn-1}=-\omega$). Physically we can think of the first statement as telling us that a quantum of energy $\hbar \omega$ was absorbed and the rotor made a transition from $n$ to $n+1$. Likewise, in the second case, the rotor lost the energy in going from $n$ to $n-1$ equal to $-\hbar \omega$. The frequency $\omega$ will depend on the label $n$ so that $\omega = (H_{n+1n+1}-H_{nn})/\hbar$, however, in order to simplify the notation, we will omit this dependence throughout.

Therefore, the only non-zero elements of the operators $\sin\phi$ and $\cos\phi$ are the ones whose indices are adjacent. Here $n$ ranges from $-\infty$ to $+\infty$ including $n=0$. There will therefore be a two-fold degeneracy in energy $E_n=E_{-n}$ other than for $n=0$. 

We are now in a position to evaluate the elements of both the angular momentum and the trig functions of the angle. Revisiting the commutator equations we have
\begin{eqnarray}
L_{nn}(\sin\phi)_{nn+1}  - L_{n+1n+1}(\sin \phi)_{n+1n}= - i\hbar (\cos \phi)_{nn+1} \\
L_{nn}(\cos\phi)_{nn+1}  -  L_{n+1n+1}(\cos \phi)_{n+1n}= i\hbar (\sin \phi)_{nn+1}
\end{eqnarray}
and similarly for the elements $nn-1$. The solution to these simultaneous equation is $L_{nn}=n\hbar$ and $(\cos \phi)_{nm} = (\delta_{nm+1}+\delta_{nm-1})/2$ and $(\sin \phi)_{nm} = (\delta_{nm+1} - \delta_{nm-1})/2i$. Note briefly that the $x$ and $y$ operators do commute as they should. Note also that another way in which we could have computed the angular momentum operator is via the relationship $L=xp_y - yp_x$, where $p_x = \mu \dot x$ and $p_y = \mu \dot y$. The matrix elements of the momenta are $(p_x)_{nm} = i\mu \omega_{nm}x_{nm}$ and $(p_y)_{nm} = i\mu \omega_{nm}y_{nm}$. It is straightforward to confirm that, as expected, $[x,p_x]=i\hbar$ and $[y,p_y]=i\hbar$, while $[p_x,p_y]=0$. This shows a further consistency in the quantum-classical correspondence.  

The bottom line is that we have obtained the spectrum of the Hamiltonian $H_{nn}= \hbar^2 n^2/(2\mu r^2)$ without ever knowing (or needing to know) the eigenvectors. The temporal dependence is given by the usual factor $e^{-iH_{nn}t/\hbar}$. Normally, we would proceed in the Schr\"odinger picture to use the wrong commutator $[\phi,L]=i\hbar$ to infer that $L=-\hbar \partial/\partial \phi$ and the time independent Schr\"odinger equation would then tell us that 
\begin{equation}
-\frac{\hbar^2}{2I}{\partial^2\over{\partial \phi^2}}\psi (\phi) = E \psi (\phi) \; .
\end{equation}
The solutions are of the form $\psi = e^{i n \phi}$ such that $n$ is an integer due to the single valued requirement that $\psi(\phi+2\pi) = \psi (\phi)$. Furthermore, the states could be normalised so that $\int_0^{2\pi} |\psi|^2 d\phi = 1$. This is consistent with the fact that $(\cos \phi)_{nm} = \int_0^{2\pi} \cos\phi |\psi|^2 d\phi$ and gives us the same value of the elements as we obtained by the Heisenberg method. The same is true for other operator elements we obtained above. 

It may seem that the Schr\"odinger representation leads us to the same result, only more economically. However, the economy is there because we took the short-cut via the wrong commutation relations and then used the analogy with the position and momentum coordinates. In any case, given that the two methods yield the same spectrum, does this mean that we do not - at least in principle - need the eigenvectors? For Heisenberg, it was sufficient to obtain the spectra because that was what the experiments were telling him at the time. He needed to explain various lines that were observed to be emitted by the atoms subject to spectroscopy.

But what if the atoms did not start in the ground state or, for that matter, in any of the eigenstates of energy? If we had a superposition of energy eigenstates initially, would we then need states in order to calculate the values of various probabilities for different measurement outcomes at a later stage?  The surprizing answer is no. 

Suppose our initial state is a superposition of two energy eigenstates $|\psi\rangle = (|1\rangle +|2\rangle)\sqrt{2}$. We then want to know the expected value of angular momentum at time $t$. This is given by the Born rule:
\begin{equation}
\langle L\rangle (t) = \langle \psi| L(t)|\psi\rangle 
\end{equation}
which would suggest to us that we need the state. The initial state is always an outcome of our experimental preparation or of the circumstances found at the start of the experiment (e.g., the atoms at room temperature tend to be in the ground state because the gap to the first excited state is much larger than the corresponding $kT$). 

But, and this is the main point of this paper, we need not express the Born rule in terms of states. We could always say that we start in a specific superposition of eigenstates $E$. This way we can stay spiritually true to the Heisenberg methodology and work with operators only since what we are interested in ultimately is the expected value of another observable, say $O$. We need to specify this operators commutation relations with the energy operator, $[O,H]=A$, as we did above for the trig functions of the angle. From these commutation relations we can find the elements of $O$ in the basis in which the energy is diagonal, i.e., $O_{nm}$. Suppose that we start with an equal superposition of $n=1$ and $n=2$ states of $H$. What is the expected value of $O(t)$ at time $t$? The answer is:
\begin{equation}
\langle O\rangle (t) = \langle \psi| O(t)|\psi\rangle = \frac{1}{2}(O_{11}(t)+ O_{12}(t)+O_{21}(t)+O_{22}(t))
\end{equation}
and the final expression contains only the operator elements. 

As an example, let us say that we start from an equal superposition of the rotor angular momentum eigenstates $n=1$ and $n=2$. Suppose that we want to know the expected value of the position operator at a later time $t$. This is, according to the previous formula, given by 
\begin{equation}
\langle x\rangle (t) = \frac{1}{2}(x_{11}(t)+ x_{12}(t)+x_{21}(t)+x_{22}(t))=\frac{r}{2} (e^{i\omega t} + e^{-i\omega t}) = r\cos \omega t =r\cos \phi \; .
\end{equation}
This is exactly what is expected from the dynamics of the classical rotor, though it should be noted that the value of $\omega$ is dependent on $n$. Therefore, the Born rule can be phrased in a way that we do not need to know the form of the eigenstates. It is sufficient to know the commutation rules between the relevant operators and rely on the fact that the Heisenberg's equations of motion (i.e., the Hamilton's equations for operators) fully specify the dynamics of the system.

In summary, we have shown how to calculate the only observationally relevant quantum mechanical quantity, the expected value of an observable given the initial state, by using the operators only and without any need to know specifically the wave-functions of the system. We were interested in diagonalising the Hamiltonian, but there is nothing special about working in the energy eigen-bases. We could have worked in the basis of an arbitrary operator. 

This makes the analogy between quantum and classical mechanics more transparent. By specifying the initial state through the corresponding value(s) of a specific operator, as well as the relevant commutation relations, we are then able to compute all the other physically relevant future values. Of course, the key difference is that, classically, the initial condition is given by the initial values of both the position and the momentum (or other canonical variables), while, quantumly, this clearly cannot be done due to their complementary nature. The point, however, is that in both classical and quantum physics the initial conditions are comprised of real numbers (i.e., commuting entities). In the quantum case this can be only one number (an integer), i.e. the eigenstate of a suitably specified observable, or, if the state is mixed, it implies specifying the relevant integers and the corresponding probabilities (real numbers). This simply reflects the information we have about the state of the system at the beginning of the experiment.

\textit{Acknowledgments}: Vlatko Vedral's research is supported by the Gordon and Betty Moore and Templeton Foundations.

\end{document}